\documentclass[letterpaper]{article}
\usepackage{natbib,alifeconf}

\title{Four Classes of Morphogenetic Collective Systems}
\author{Hiroki Sayama\\
\mbox{}\\
Collective Dynamics of Complex Systems Research Group\\
Binghamton University, State University of New York, Binghamton, NY 13902-6000, USA\\
sayama@binghamton.edu}

\begin{document}
\maketitle

\begin{abstract}
We studied the roles of {\em morphogenetic principles}---heterogeneity
of components, dynamic differentiation/re-differentiation of
components, and local information sharing among components---in the
self-organization of morphogenetic collective systems. By
incrementally introducing these principles to collectives, we defined
four distinct classes of morphogenetic collective systems. Monte Carlo
simulations were conducted using an extended version of the Swarm
Chemistry model that was equipped with dynamic
differentiation/re-differentiation and local information sharing
capabilities. Self-organization of swarms was characterized by several
kinetic and topological measurements, the latter of which were
facilitated by a newly developed network-based method. Results of
simulations revealed that, while heterogeneity of components had a
strong impact on the structure and behavior of the swarms, dynamic
differentiation/re-differentiation of components and local information
sharing helped the swarms maintain spatially adjacent, coherent
organization.
\end{abstract}

\section{Introduction}

Self-organizing behaviors of biological collectives have been subject
to many scientific inquiries \citep{1,2,3,4,5,6}. They have also been
enthusiastically applied to engineering problem solving as a new
paradigm and methodology for decentralized, distributed problem
solving by collaborative artificial agents
\citep{7,8,9,10,11,12,13,14,16,15}. Typical assumptions made in the
existing literature are that the system components interact with each
other mostly locally and make decisions at individual levels, which
eventually leads to the emergence of non-trivial (and potentially
useful) macroscopic behaviors. Those models have been successful in
reproducing various self-organizing patterns and adaptive
functionalities.

However, theoretical models used in earlier studies were predominantly
focused on homogeneous physical collectives and animal
populations. While their simplicity is by itself a virtue in some
regard \citep{6}, they are often too simple to capture more complex
phenomena seen in real-world biological collectives, such as
multi-cellular organisms' morphogenesis and physiology, termite colony
building and maintenance, and growth and self-organization of human
social systems. These systems operate with highly sophisticated
within-system regulation mechanisms, or ``programs''
\citep{16,15}. What are common among those real-world complex
biological collectives are {\em heterogeneity of components}, {\em
  dynamic differentiation/re-differentiation of components} (i.e.,
dynamic switching of component types/roles), and {\em local
  information sharing among components} that influence their
differentiation. These {\em morphogenetic principles} are often absent
in earlier models of biological collective behaviors, and they have
not been fully utilized in engineering applications either. Here we
define a {\em morphogenetic collective system} as a system made of a
large number of components that self-organize to form nontrivial
structures and behaviors using these morphogenetic principles.

There is growing literature of analytical and numerical studies on
self-organizing behavior of biological collectives
\citep{6,17,18,19,20,21,25,22,23,24,26,27,28,29}. It is repeatedly
reported that there are a small number of universal classes of
collective behaviors, such as disordered, highly ordered, rotational,
critical, and jamming patterns, as well as various forms of phase
transitions between those classes \citep{6}. Most of these results are
based on the assumption that collectives are homogeneous in terms of
their components' kinetic and behavioral properties. Within-population
variations are rarely considered in those theoretical models.

There are some studies that considered the effects of heterogeneity
within biological collectives. Graves et al.~studied mixed-species
bird flocks in Amazonia and created a computational model of them
\citep{30}. More recently, Couzin et al.~studied self-sorting of a fish
school caused by physical variations among individuals \citep{31}. His
group also studied collective decisions of a swarm influenced by a
small number of informed individuals, or leaders \citep{32}. We also
proposed the Swarm Chemistry model \citep{sc,cim,gecco,mechapter} to
explore self-organization of swarms made of kinetically distinct types
of particles. However, these studies still assumed that
within-population variations are variations of fixed individual
properties. None of them considered a more sophisticated form of
dynamic, adaptive changes of behavioral rules of individuals within a
population, potentially through local communication and information
sharing.

When one looks at real-world biological collectives, there are a
number of examples where more complex forms of heterogeneous
collectives produce highly intricate patterns and behaviors that look
almost self-evidently ``designed'' by someone or something
\citep{16,15,33}. Such instances can be found at every scale in
biology. For example, Ben-Jacob et al.~reported very complex,
heterogeneous, even intelligent, information processing and motion
control taking place in bacteria societies \citep{2,34,35}. Social
insects are another well-studied example, where dynamic switching of
different roles driven by local information sharing realizes highly
efficient division of labor \citep{4,8,36,37,38}. Among the most
interesting and complex examples are the incredibly sophisticated,
large-scale mounds built and maintained by termites
\citep{39,33,40}. A termite mound operates as if it were a carefully
designed and fully integrated physiological system, inside which
individual termites ``differentiate'' to play various different
roles. The complexity of such real-world biological collectives were
not fully captured in the earlier literature of collective behaviors
mentioned above.

\section{Four Classes of Morphogenetic Collective Systems}

\begin{figure*}[t]
\centering
\includegraphics[width=0.8\textwidth]{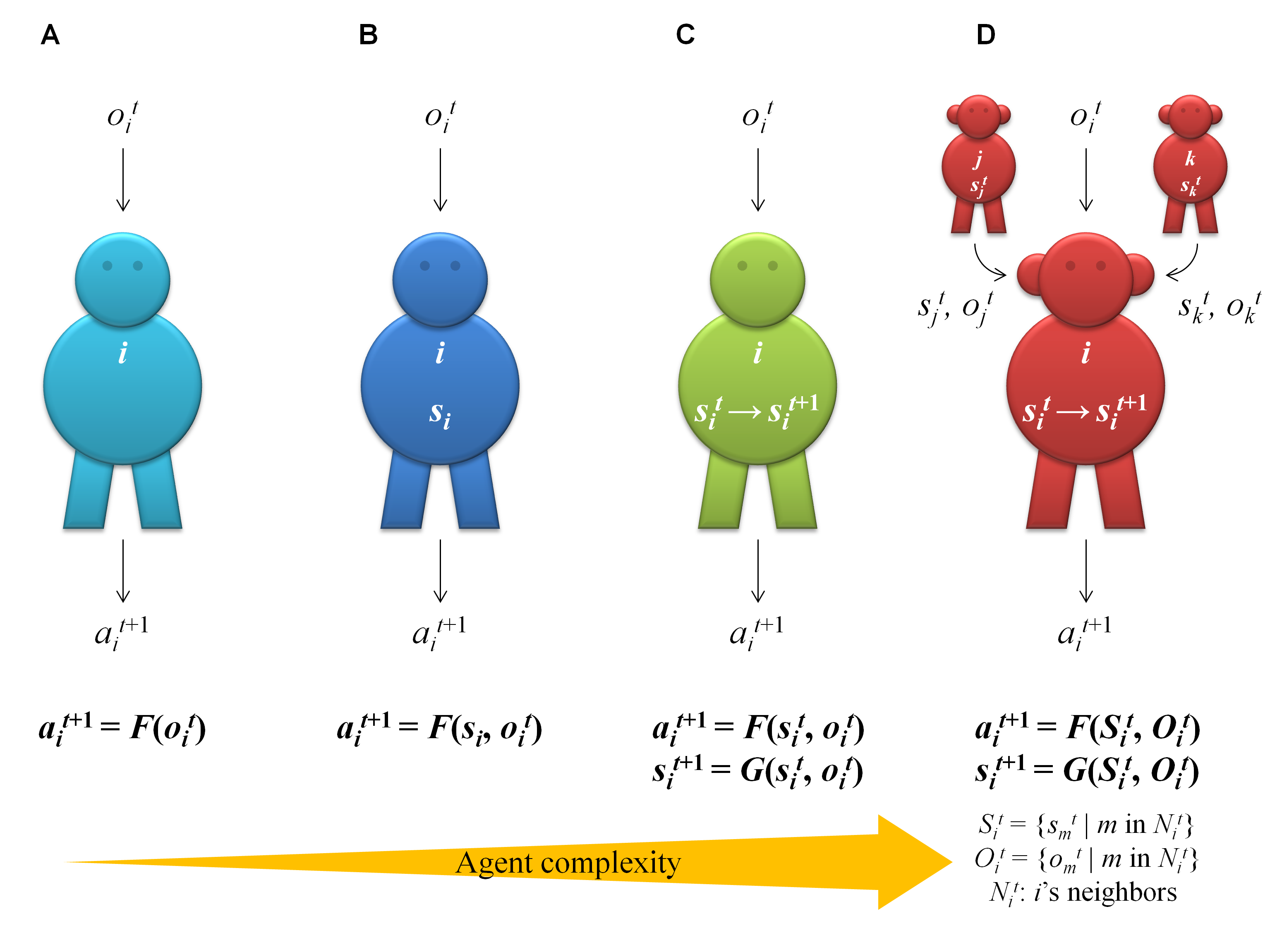}
\caption{Four classes of morphogenetic collective systems proposed in
  this paper. Variables $s_i$, $o_i$ and $a_i$ represent the internal
  state of agent $i$, the observation it receives from the
  environment, and the corresponding action it takes (e.g.,
  acceleration), respectively. A: Homogeneous collectives. Agents'
  behaviors are determined by a function of observations,
  $F(\circ)$. B: Heterogeneous collectives. Each agent has its own
  static state ($s_i$), and $F$ takes $s_i$ as an additional argument
  as well as $o_i$. C: Heterogeneous collectives with dynamic
  differentiation/re-differentiation. Agents' states can dynamically
  change according to another function, $G(\circ)$. D: Heterogeneous
  collectives with dynamic differentiation/ re-differentiation and
  local information sharing. Arguments of the functions are sets of
  $s_i$ and $o_i$ within the agent's neighborhood, which represents
  the local information sharing.}
\label{fourclasses}
\end{figure*}

As briefly reviewed above, the dynamics and capabilities of different
types of morphogenetic collective systems are yet to be fully
understood. Recent increase of studies that incorporate at least part
of the morphogenetic principles indicates the promising nature of this
direction of research. In order to systematically study the effects of
each of the morphogenetic principles, here we propose the following
four distinct classes of morphogenetic collective systems, which are
obtained by incrementally introducing morphogenetic principles to
agents' behavioral and communication capabilities
(Fig.~\ref{fourclasses}):
\begin{enumerate}
\item[A.] {\em Homogeneous collectives}, where agents' behaviors are
  determined by a globally defined, uniformly applicable function of
  observations.
\item[B.] {\em Heterogeneous collectives}, where an agent's behavior
  is determined by a function of observations specified by the agent's
  static state or type.
\item[C.] {\em Heterogeneous collectives with dynamic differentiation/
  re-differentiation}, where the heterogeneity of agent behaviors is
  created and dynamically maintained by transitions of agents'
  internal states. Agents' state transitions are also determined by a
  function of observations and states.
\item [D.] {\em Heterogeneous collectives with dynamic
  differentiation/ re-differentiation and local information sharing},
  where agents can share information (internal states and their
  observations) with local neighbors, in addition to all the above
  capabilities.\footnote{Here we consider information sharing as
    explicit signal transmission among agents to share externally
    unobservable information about their internal states and
    observations with local neighbors. This should not be confused
    with kinetically transferred information theoretic waves
    propagating in swarms that were studied in recent literature
    \citep{infocascade}.}

\end{enumerate}
Our rationale in defining these four classes is that each
morphogenetic principle requires the precedent one. Namely,
differentiation/re-differentiation requires multiple states or types
of components (heterogeneity), and information sharing would make
sense only if agents could change their behaviors according to it
(differentiation/re-differentiation). We thus argue that these four
classes should represent a natural, straightforward hierarchy of
morphogenetic collective systems, arranged in the ascending order of
their organizational complexity.

In this classification, Class A includes traditional collective
behavior models, including Boids \citep{1} and its variants that use
homogeneous swarms. More recent models that use heterogeneous swarms
\citep{31,32,sc,gecco} belong to Class B. Examples that may be close
to those of Class C are the variants of Swarm Chemistry
\citep{cim,ieeealife,52} that implemented stochastic
differentiation/re-differentiation of agents, though it was not driven
by any state-transition function and thus not dynamical or
adaptive. Finally, real biological/social morphogenetic systems---such
as embryogenesis of multicellular organisms, colonies of eusocial
insects and cities in human civilizations---are most likely in Class
D. Several models proposed in Morphogenetic Engineering \citep{mebook}
also belong here. However, we are not aware of any well established
model of swarm-based collective behaviors that belong to this class.

\section{Model: Morphogenetic Swarm Chemistry}

To study qualitative and quantitative differences in possible
morphologies and behaviors between the four classes, we have developed
a mathematical model of morphogenetic collective systems by
implementing new rule-based state transition and local information
sharing capabilities in the Swarm Chemistry model
\citep{cim,mechapter}. Swarm Chemistry is naturally suitable for this
research task because it already can represent both homogeneous and
heterogeneous collective systems in its model framework.

In the extended model, the design of a swarm is specified in three
parts: a recipe $\cal R$, a preference weight matrix $U$, and a local
information sharing coefficient $w$. Definitions of these parts are as
follows:
\begin{description}
\item[Recipe $\cal R$:] A list of different kinetic parameter settings
  for multiple swarm states. Each entry in a recipe is composed of a
  relative frequency of a particular state within the swarm and its
  kinetic parameter settings (e.g., local perception range, normal
  speed, strengths of kinetic forces, etc.). This part of the
  specification is the same as in our earlier studies \citep{sc,cim}.
\item[Preference weight matrix $U$:] A $n \times (n+5)$ rectangular
  real-valued random matrix where $n$ is the length of the recipe,
  i.e., the number of possible states of agents in a swarm. Its
  contents represent how each observation component (defined later)
  affects the agent's preference for a particular state choice.
\item[Local information sharing coefficient $w$:] A real number in
  $[0,1]$, which determines how much information about the neighbors'
  states and observations are to be shared and incorporated into the
  agent's own state transition.
\end{description}

Each agent in a swarm in this extended model has its own state $s_i$,
in addition to position $x_i$ and velocity $v_i$. The movement and
state transition of agents is simulated as follows:
\begin{description}
\item[Step 1.] An agent computes its action (acceleration) based on
  its neighbors' relative positions and velocities using the standard
  Swarm Chemistry simulation algorithm \citep{sc}.
\item[Step 2.] Before making any actual movement, the agent also
  computes an $(n+5)$-dimensional {\em observation vector} $o_i$ that
  summarizes the situation the agent is in. More details of this
  vector will be discussed later.
\item[Step 3.] Once all the agents computed their actions and
  observation vectors, each agent updates its velocity according to
  the computed acceleration, and then moves using the updated
  velocity.
\item[Step 4.] The agent computes a {\em state preference vector} $u_i
  = (1-w)Uo_i + wU\langle o\rangle_i$, where $\langle o\rangle_i$ is
  the average of observation vectors of other agents in the local
  neighborhood. If there are no other agents found in the
  neighborhood, $u_i = Uo_i$ regardless of $w$.
\item[Step 5.] The agent checks if the $s_i$-th component of $u_i$,
  $u_i(s_i)$, is negative. If this is the case, it means that its
  current state is not preferable in the current situation, and
  therefore the agent attempts, with probability $1 - \exp(u_i(s_i))$,
  to choose a new state. A new state will be chosen as the next value
  of $s_i$, via a roulette selection where $\exp(u(s))$ is used as the
  selection weight for state $s$.
\end{description}

The observation vector $o_i$ plays an important role in this
model. While there are many choices for how to construct an
observation vector, we used the following definition as an initial
step of our investigation. The first $n$ components of $o_i$ are all
0's except for the $s_i$-th component that is set to 1. This part
embeds the information about the agent's current state into the vector
so that it can be superposed and averaged with other agents'
states. The remaining five components of $o_i$ are as follows:
\begin{itemize}
\item $|\langle x \rangle_i - x_i|^2/R_i^2$: Square of the relative
  distance from the average position of other agents in the
  neighborhood ($\langle x \rangle_i$). $R_i$ is the perception range
  in the parameter set the agent is currently using. If there are no
  other agents nearby, this is set to 0.
\item $v_i^2/v_{i n}^2$: Square of the ratio of the agent's current
  velocity and the normal velocity ($v_{i n}$) in the parameter set
  the agent is currently using.
\item $\langle v \rangle^2/v_{i n}^2$: Square of the ratio of the
  neighbors' average velocity ($\langle v \rangle$) and the normal
  velocity in the parameter set the agent is currently using. If there
  are no other agents nearby, this is set to 0.
\item $|\langle v \rangle_i - v_i|^2/v_{i n}^2$: Square of the
  relative difference in velocity between the agent and its
  neighbors. If there are no other agents nearby, this is set to 0.
\item $1$: A constant term.
\end{itemize}

Although it is a highly constrained, mathematically stylized
formulation, this extended Swarm Chemistry model can still describe a
wide variety of morphogenetic collective systems, including the four
classes proposed in this paper. Specifically, including only one
parameter set in $\cal R$ and letting $U=0$ and $w=0$ will make a
Class A swarm. Including multiple parameter sets in $\cal R$ with
$U=0$ and $w=0$ will make a Class B swarm. A Class C swarm will be
obtained by additionally adopting a non-zero matrix for $U$ while
$w=0$. Finally, a Class D swarm will be obtained by adopting non-zero
$U$ and $w$.

\section{Experiments}

We conducted systematic Monte Carlo simulations to see if there were
any significant differences in the dynamics of morphogenetic
collective systems among the four classes. Specifications of swarm
designs were configured as follows:
\begin{itemize}
\item The number of agents was fixed to 300 for all cases.
\item The number of possible agent states $n$ was set to 1 for Class A
  swarms, or otherwise $\xi + 2$ where $\xi$ is a random integer
  sampled from a Poisson distribution with mean 1.
\item Recipe $\cal R$ was created by sampling each kinetic parameter's
  value from a uniform distribution between 0 and the parameter's
  maximum value defined in \citep{sc}. The relative frequencies of
  different parameter settings were determined by randomly dividing
  300 into $n$ portions.
\item Preference weight matrix $U$ was set to all 0's for Class A and
  B swarms. For Class C and D swarms, each component of $U$ was
  sampled from a normal distribution with mean 0 and standard
  deviation 0.1.
\item Local information sharing coefficient $w$ was set to 0 for
  Class A, B and C swarms, while it was sampled from a uniform
  distribution between 0 and 1 for Class D swarms.
\end{itemize}

Five hundred independent simulation runs were conducted for each
class. Each run was initialized with 300 agents whose positions were
randomly distributed in a two-dimensional $300 \times 300$ (in
arbitrary unit) square area and whose velocities were set to 0, and
then simulated for 400 time steps. In each time step during the second
half of the simulation ($t=201-400$), the following properties were
measured from the agents' positions and velocities:
\begin{itemize}
\item Average speed of the swarm as a whole ($|\langle v \rangle|$,
  where $\langle \ldots \rangle$ denotes an average over all the
  agents hereafter unless noted otherwise).
\item Average of absolute speed of agents ($\langle |v| \rangle$).
\item Average angular velocity of the swarm as a whole ($\langle (x -
  \langle x \rangle) \times v / |x - \langle x \rangle|^2\rangle$).
\item Average distance of agents from center of mass ($\langle |x -
  \langle x \rangle| \rangle$).
\item Average pairwise distance ($\langle |x_1 - x_2| \rangle$, where
  $x_1$ and $x_2$ are positions of two randomly sampled agents; the
  average in this case was over 10,000 sampled pairs).
\end{itemize}

\begin{figure}
\centering
\includegraphics[width=\columnwidth]{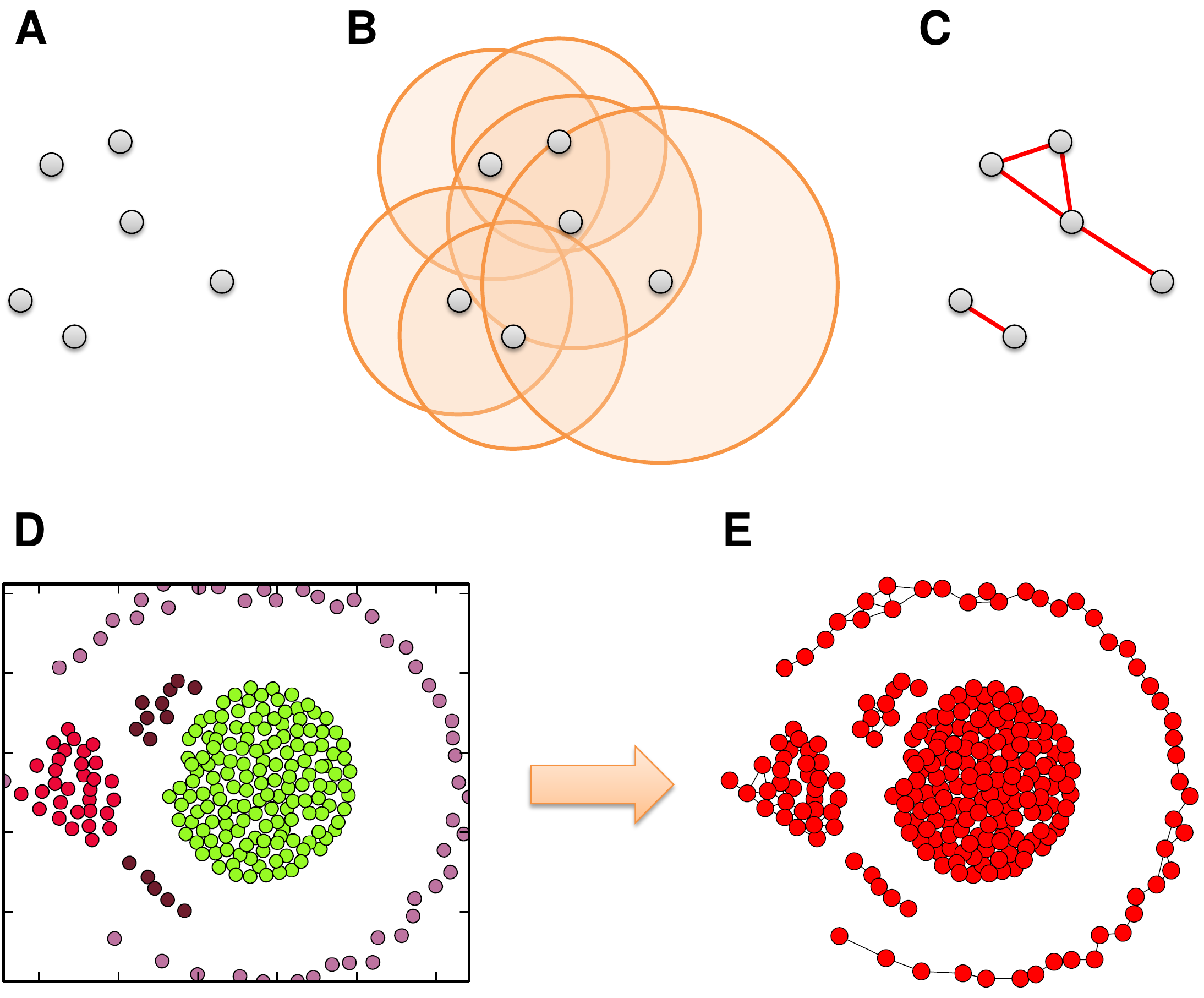}
\caption{Construction of networks from swarms. Top: Illustration of
  the algorithm. A: Spatially distributed agents. B: Ranges of
  neighbor recognition drawn around the agents. The radius of each
  range is $\alpha d_i$ (see text). C: Resulting network where pairs
  of agents that mutually recognize each other as neighbors are
  connected. Bottom: An example of network construction from a
  simulated swarm. D: Original swarm snapshot. E: Network constructed
  from agent positions in D.}
\label{cluster-graph}
\end{figure}

In addition, topological properties of the swarm morphology were also
measured. For this task, we constructed a network for each swarm at
each time step during the measurement period by connecting agents that
were spatially close to each other. More specifically, we first
measured a characteristic ``neighbor'' distance $d_i$ for each agent
by calculating the mean of its distances to its $k$ nearest
neighbors. The agent then recognized all other agents within distance
$\alpha d_i$ as its ``neighbors''. Once this was done for all agents,
the pairs of agents that mutually recognized each other as
``neighbors'' were actually connected. We used $k=2$ and $\alpha =
1.7$, which were empirically chosen to approximate cluster structures
recognized visually by human experimenters. Figure \ref{cluster-graph}
shows examples of this network construction process.

For each swarm converted to a network, we measured the following
topological properties:
\begin{itemize}
\item Number of connected components.
\item Average size of connected components.
\item Homogeneity of sizes of connected components. This was measured
  by the normalized entropy in the distribution of sizes of connected
  components. If there was only one connected component, this was set
  to 1.
\item Size of the largest connected component.
\item Average size of connected components smaller than the largest
  one. If there was only one connected component, this was set to 0.
\item Average clustering coefficient.
\item Link density.
\end{itemize}

Altogether, we obtained time series of $5+7=12$ measurements from each
simulation run. These time series were further summarized by
calculating their respective mean and standard deviation over time. As
a result, the structure and behavior of each swarm was characterized
by $2 \times 12 = 24$ outcome variables. Finally, a visual image of
the swarm at the end of the simulation ($t=400$) was also recorded for
visual inspection of the swarm's topology.

The simulator was implemented in Python using NetworkX
\citep{networkx} and PyCX \citep{pycx}. The code is available upon
request.

\section{Results}

Simulation results are summarized in Table \ref{resulttable}, where
medians of 24 outcome variables were shown for each class together
with $p$-values obtained using the Kruskal-Wallis median test between
the four classes.\footnote{Visual snapshots of the swarms' final
  configurations for each of the four classes are also available
  online at
  http://bingweb.binghamton.edu/$\tilde{~}$sayama/SwarmChemistry/.}
Statistically significant differences were detected for most of the
outcome variables, especially for topological outcome variables, which
demonstrates the effectiveness of our network-based topology
characterization method. Most of the variables with statistical
significance showed clear differences between Class A (homogeneous)
and other (heterogeneous) swarms. The temporal mean of the average
clustering coefficient was the only outcome variable that did not show
statistically significant differences between the classes. This metric
may be primarily determined by constraints built in our simulation or
network construction algorithms.

\begin{table*}[t]
\caption{Comparison of medians of 24 outcome variables between four
  classes of morphogenetic collective systems. The $p$-values of the
  Kruskal-Wallis median test were shown in the rightmost column (*:
  $p<0.01$, **: $p<0.001$, ***: $p<0.0001$). Variables with
  statistically significant differences were highlighted in color
  (yellow: high, cyan: low). The intensity of color is adjusted
  according to the $p$-value of the variable and the distance from
  overall average. The top half shows temporal means of the 12
  measurements while the bottom half shows temporal standard
  deviations of the 12 measurements.}
~\\ \centering
\includegraphics[width=0.845\textwidth]{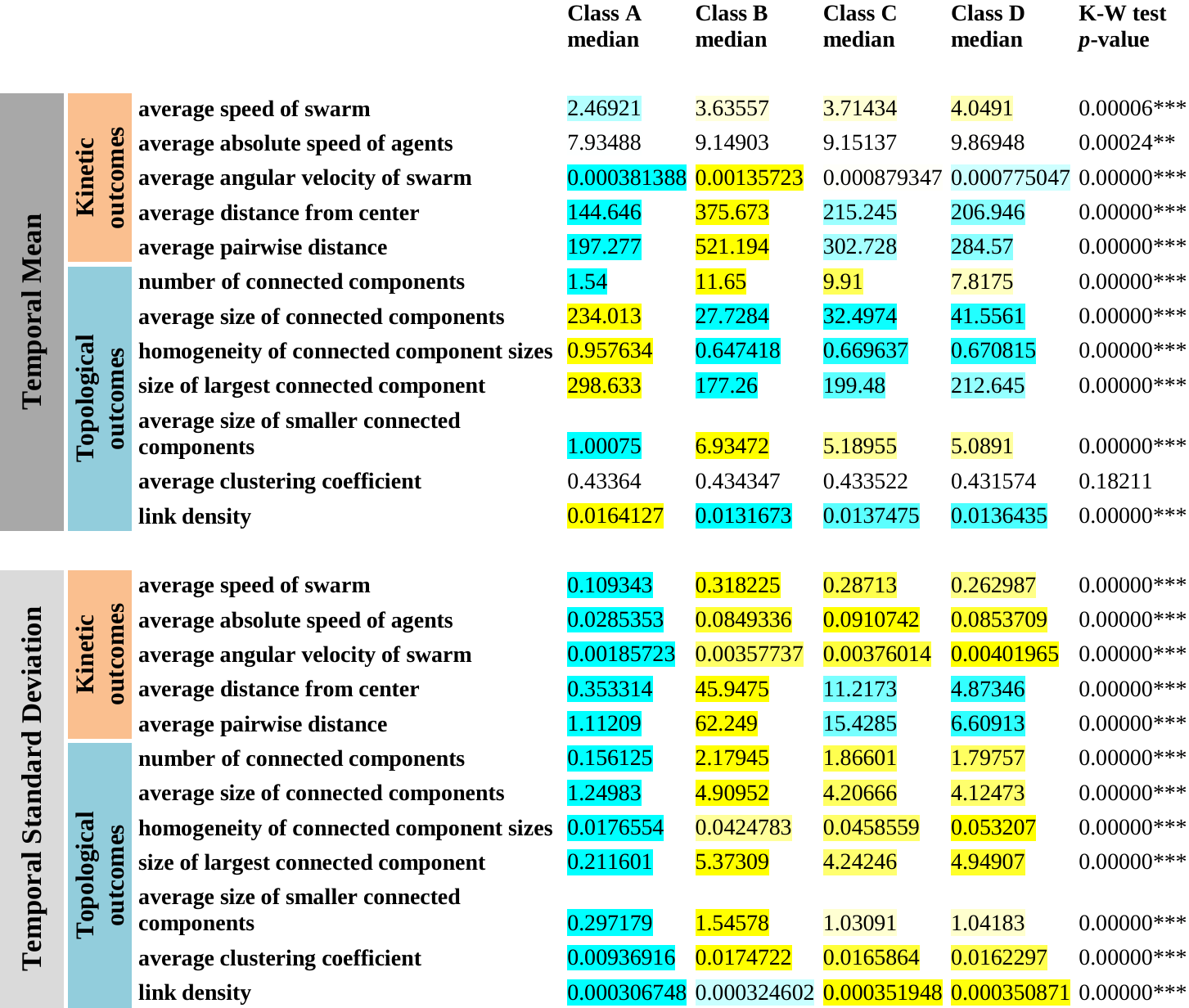}
\label{resulttable}
\end{table*}

We found several notable patterns in Table \ref{resulttable}. First,
the medians of Class A swarms consistently took the lowest values
among the four classes with regard to the temporal standard deviations
of measurements (lower half of Table \ref{resulttable}). This
indicates that the interactions between different types of agents
helped produce dynamic behaviors, causing temporal fluctuations of
their macroscopic properties. Second, there were clear differences
between Class B (with no state transitions) and Classes C \& D (with
state transitions) regarding temporal standard deviations of the
average distance of agents from the center of mass and the average
pairwise distance. This is likely due to the fact that, while Class B
swarms tend to disperse into smaller clusters easily, agents of Class
C \& D swarms can stay together and maintain spatially adjacent,
coherent organization more often, because adaptive state transitions
help initially incompatible agents assimilate into kinetically
compatible types.

Finally, we noticed a consistent trend in a number of the outcome
variables that the properties of Class C \& D swarms sat somewhere in
between those in Class A and Class B. This could also be understood in
that dynamic state transition, possibly driven by local information
sharing, has enabled swarms to adaptively achieve coherence in their
structures and behaviors. However, the results also indicate that the
swarms in Classes C \& D did not simply turn into a more homogeneous
state like Class A ones, because nearly all the topological outcome
variables showed significant differences between Class A and Classes C
\& D. Therefore, this apparent trend found in morphogenetic collective
systems in Classes C \& D must be understood not as simple
homogenization but as an emergent shift of higher-level system
properties.

\section{Conclusions}

In this paper, we proposed a classification of morphogenetic
collective systems based on the absence or presence of three
morphogenetic principles: heterogeneity of components, dynamic
differentiation/re-differentiation of components, and local
information sharing among components. Monte Carlo simulations with an
extended morphogenetic Swarm Chemistry model demonstrated that, while
heterogeneity of components had a strong impact by itself on the
structure and behavior of the swarms, the other two morphogenetic
principles, i.e., dynamic differentiation/re-differentiation of
components and local information sharing, greatly contributed to the
maintenance of spatially adjacent, coherent organization of
swarms. Interestingly, many outcome measurements of Class C and D
swarms fell somewhere in between those of Class A and Class B. This
result indicates that the dynamic, adaptive state transition of
components possibly driven by their mutual information sharing is
playing an essential role in achieving structural and functional
integration of biological and social collectives.

The present study has several fundamental limitations. First, the
dynamics of swarms were explored only through a limited number of
random parameter sampling and analyzed only using simple median
comparisons. It was a reasonable first step of exploration when
nothing was known about the model, but this approach would not be able
to characterize behavioral diversity and richness of each class of
systems or to discover non-trivial behaviors that would be
statistically rare but unique and interesting. To fully explore and
understand the limit of dynamical diversity of each class, more
sophisticated evolutionary or other population-based search methods
should be conducted. Now that we have 24 outcome metrics defined, we
can try evolving morphogenetic collective systems toward a certain
area in this metric space, to examine how closely the swarms of each
class can achieve the target properties.

Second, the model we used in this study (morphogenetic Swarm
Chemistry) was developed using somewhat arbitrary design
decisions. The behavioral rules were exactly the same as those used in
previous swarm models, which were limited to agents' acceleration
only. The state transition functions were defined in a rather simple
and linear fashion using a product of a preferential weight matrix and
an observation vector. Moreover, the variables included in the
observation vector were chosen without substantial justification. We
will need to consider adopting more open-ended, nonlinear forms of
representations for observations, state transitions and agent
behaviors. We plan to apply genetic programming \citep{49} or other
symbolic evolutionary search methods \citep{50,51} to overcome this
limitation in the future.

Finally, the scope of the experiments was limited only to ``free''
self-organization of collectives without any stimuli or
constraints. Exposing swarms to such external conditions and measuring
their adaptive responses will further clarify the functional
differences between different classes of morphogenetic collective
systems.

\section{Acknowledgments}

This material is based upon work supported by the US National Science
Foundation under Grant No.~1319152.

\footnotesize
\bibliographystyle{apalike}

\end{document}